\documentclass[acus,%        % or letter
              ]{jacow}

\makeatletter%                           % test for XeTeX where the sequence is by default eps-> pdf, jpg, png, pdf, ...
\ifboolexpr{bool{xetex}}                 % and the JACoW template provides JACpic2v3.eps and JACpic2v3.jpg which might generates errors
 {\renewcommand{\Gin@extensions}{.pdf,%
                    .png,.jpg,.bmp,.pict,.tif,.psd,.mac,.sga,.tga,.gif,%
                    .eps,.ps,%
                    }}{}
\makeatother

\ifboolexpr{bool{xetex} or bool{luatex}} % test for XeTeX/LuaTeX
 {}                                      % input encoding is utf8 by default
 {\usepackage[utf8]{inputenc}}           % switch to utf8

\usepackage[USenglish]{babel}

\ifboolexpr{bool{jacowbiblatex}}%        % if BibLaTeX is used
 {%
  \addbibresource{jacow-test.bib}
  \addbibresource{biblatex-examples.bib}
 }{}

\usepackage{subfig}
\usepackage{float}
\usepackage{multirow}
\usepackage{epstopdf}
\begin{document}

\title{Advancement in the understanding of the field and frequency dependent microwave surface resistance of niobium}

\author{M. Martinello\thanks{mmartine@fnal.gov}, S. Aderhold, S. K. Chandrasekaran, M. Checchin, A. Grassellino,  O. Melnychuk,\\ S. Posen, A. Romanenko, D.A. Sergatskov \\ Fermi National Accelerator Laboratory, Batavia, IL 60510, USA}

\maketitle
\begin{abstract}
The radio-frequency surface resistance of niobium resonators is incredibly reduced when nitrogen impurities are dissolved as interstitial in the material, conferring ultra-high Q-factors at medium values of accelerating field. This effect has been observed in both high and low temperature nitrogen treatments. As a matter of fact, the peculiar anti Q-slope observed in nitrogen doped cavities, i.e. the decreasing of the Q-factor with the increasing of the radio-frequency field, come from the decreasing of the BCS surface resistance component as a function of the field. Such peculiar behavior has been considered consequence of the interstitial nitrogen present in the niobium lattice after the doping treatment. The study here presented show the field dependence of the BCS surface resistance surface of cavities with different resonant frequencies, such as: 650 MHz, 1.3 GHz, 2.6 GHz and 3.9 GHz, and processed with different state-of-the-art surface treatments. These findings show for the first time that the anti Q-slope might be seen at high frequency even for clean Niobium cavities, revealing useful suggestion on the physics underneath the anti Q-slope effect. 
\end{abstract}

\section{Introduction}

Superconducting Radio-Frequency (SRF) cavities are key components of modern particle accelerators. For continuous wave (CW) accelerators it is extremely important to maximize the cavity Q-factor in order to lower the power dissipated in the cavity walls and, therefore, the cryogenic cost. The so-called Nitrogen-doping is a surface treatment capable to dramatically improve the SRF performance, increasing the Q-factor by a factor of three at medium values of accelerating field, i.e. around $E_{acc}=16$ MV/m\cite{Grassellino_SUST_2013}. Peculiar signature of such a treatment is the increasing of the quality factor as a function of the accelerating field, called anti Q-slope to underline that the trend is opposite to the usual Q-slope observed at medium accelerating field, in standard treated niobium cavities.

The R$\&$D of SRF cavities in the last decades has been particularly focused in studying 1.3 GHz cavities, as of major importance for large accelerator project such as the International Linear Collider (ILC), the European X-ray Free Electron Laser (EXFEL) and the Linear Coherence Ligth Source II (LCLS-II). This is especially true for the N-doping treatment that, right after its discovery in 2013, has been implemented for the LCLS-II cryomodule production. This paper show the first systematic study of the field dependent micro-wave surface resistance variation as a function of different resonant frequencies.

The micro-wave surface resistance $R_{s}$ is intimately related with the cavity Q-factor, being $Q_{0}=G/R_{s}$, where $ G=270$ $\Omega $ is the geometrical factor, independent on material properties.
In agreement with common convention, the RF surface resistance is defined as sum of two contributions, the BCS surface resistance ($ R_{BCS} $), and the residual resistance ($ R_{res} $). Mattis and Bardeen \cite{Mattis_PR_1958} defined for the first time the BCS surface resistance contribution, which takes into account dissipation coming from thermally excited quasi-particles, in agreement with the Bardeen-Cooper-Schrieffer theory of superconductivity  \cite{Bardeen_PR_1957}.  $ R_{BCS} $ decays exponentially with the temperature and depends on several material parameters, such as: London penetration depth $\lambda_{L}$, coherence length $\xi_{0}$, energy gap $ \Delta $, critical temperature $ T_{c} $ and mean free path $\ell$. 

Decomposing the surface resistance contribution \cite{Romanenko_APL_2013}, it has been shown that the anti Q-slope of N-doped cavities is consequence of the decreasing of the BCS surface resistance components as a function of the accelerating field. Recent studies has also suggested that this enhancement of the superconductivity at medium field,  may be due to non-equilibrium distribution of quasi-particles \cite{Martinello_APL_2016}. Findings reported later on in this paper will corroborate further such a hypothesis.

\section{Experimental Procedure}

The different surface resistance contributions ($ R_{BCS} $ and $ R_{res} $) has been calculated as a function of the accelerating field, for elliptical niobium cavities resonating at 650 MHz, 1.3, 2.6 and 3.9 GHz. A summary table of the surface treatment studied for each resonant frequency is shown in Table 1. 

\begin{table*}[t]
   \centering
   \caption{Summary of the frequency studied for each surface treatment.}
   \begin{tabular}{cccccc}
       \toprule
            & 650 MHz & 1.3 GHz & 2.6 GHz & 3.9 GHz \\
       \midrule
           BCP               &   & X &   & X &\\
           120 C baking      & X & X & X & X &\\
           2$/$6 N-doping    & X & X & X & X &\\
       \bottomrule
   \end{tabular}
   \label{tab1}
\end{table*}

As previously mentioned, the BCS surface resistance exponentially decreases as a function of the temperature. Therefore, in case of low frequencies cavities, such as 650 MHz and 1.3 GHz, the $ R_{BCS} $ contribution at $T= 1.5$ K becomes negligible compared with the residual resistance $R_{res}$, therefore $R_{S}(1.5 K) \simeq R_{res}$ and: 

\begin{equation}
R_{BCS}(2 K)=R_{s}(2 K)-R_{res} \, \text{.}
\end{equation}

In case of high frequency cavities $ R_{BCS} $ at $T= 1.5$ is not negligible instead, and in this case $R_{res}$ has to be calculated from $R_{S}(T)$ data interpolation. Since the residual resistance is independent on temperature, by fitting $R_{S}(T)$ acquired during the cavity cooldown, in a data range between 2 K and 1.5 K, it is possible to calculate at which value $R_{S}$ assets at very low temperatures, i.e. the residual resistance $R_{res}$. In first approximation the surface resistance follows the following trend:

\begin{equation}
R_{S}(T)=\dfrac{A\omega^2}{T}e^{-\frac{\Delta}{kT}}+R_{res} \, \text{,}
\end{equation}

where $A$ depends on many material parameters. The $R_{S}(T)$ curves are acquired at several values of accelerating field and interpolated using Eq. 2.  In this way it is possible to extrapolate the cavity residual resistance as a function of the accelerating field. 

The cavity mean free path was extrapolated using the SRIMP code. In this case the cavity resonance frequency as a function of temperature during the cavity warm up is acquired in order to obtain the variation of the penetration depth, $\Delta \lambda$, as a function of the temperature close to $T_{c}$ \cite{HalSurfRes}.   

The SRIMP code is used to interpolate $ \Delta \lambda $ versus temperature. The interpolation fixed parameters are: critical temperature, coherence length ($\xi_{0}=38$ nm) and London penetration depth ($\lambda_{L}=39$ nm). The parameters obtained from the interpolation are: mean free path $\ell$ and reduced energy gap (${\Delta}/{kT_{c}}$) \cite{Halbritter_KFK_1970, Martinello_APL_2016}. 

%\begin{equation}
%\begin{split}
%&R_{BCS}(2 K)=R_{s}(2 K)-R_{0} \, \text{,}\\
%&R_{fl}(B_{trap})=R_{res}(B_{trap})-R_{0} \, \text{.}
%\end{split}
%\end{equation}

%
\begin{figure}[h]
\centering
\includegraphics[scale=1]{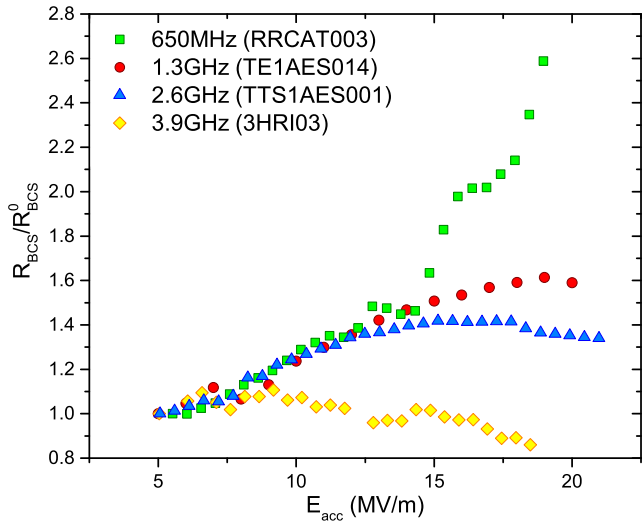}
\caption{Normalized data $ R_{BCS}/R_{BCS}^0 $ as a function of the accelerating field for  120 C baked cavities.}
\label{120}
\end{figure}
\begin{figure}[h]
\centering
\includegraphics[scale=1]{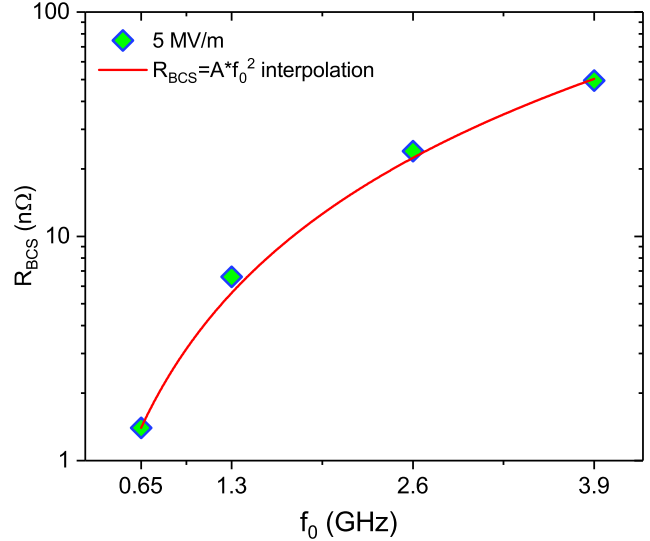}
\caption{$ R_{BCS} $ at 5 MV/m as a function of the resonance frequency for  120 C baked cavities.}
\label{LF}
\end{figure}

\section{Results and Discussion}

In order to compare the effect of each surface treatment on the BCS surface resistance, the data has been normalized for the $ R_{BCS} $ value at low field (usually between $4 MV/m$ and $5 MV/m$). 

The results of the normalized data $ R_{BCS}/R_{BCS}^0 $ for 120 C baked cavities  are shown in Figure 1. From the graph it appears that at low field the trend is very similar, but for accelerating field grater than $10 MV/m$, the field dependence starts to deviate depending on the cavity resonance frequency. Looking to the 650 MHz cavity, the increasing of $ R_{BCS} $ with the accelerating field results very steep, more than the one usually observed for the case of 1.3 GHz cavities with the same treatment. On the other hand, looking at the case of 2.6 GHz the trend seems to be more moderate, and in the case of 3.9 GHz the trend is almost reversed, showing a slight decreasing of $ R_{BCS} $ with the accelerating field. These findings are suggesting that there is a strong frequency dependence on the field dependence of the BCS surface resistance: higher frequencies are favorable to decrease the $R_{BCS}$ contribution at medium field. 

In Figure 2 the BCS surface resistance measured at 5 MV/m for these 120 C baked cavities is shown as a function of the resonance frequency. The experimental points are also interpolated with $R_{BCS}=Af_0^2$, verifying that at low field $ R_{BCS} $ does scale as the frequency square, as predicted by Mattis and Bardeen \cite{Mattis_PR_1958}.

Analogously, the normalized data $ R_{BCS}/R_{BCS}^0 $ for BCP cavities are shown in Figure 3. In this case the comparison is only between 1.3 and 3.9 GHz cavities but it is already clear how the frequency plays a major role in the $ R_{BCS} $ field dependence: differently than what happen at 1.3 GHz, at 3.9 GHz $ R_{BCS} $ decreases as a function of the accelerating field. This is the first clear proof that the BCS surface resistance can decrease as a function of the field even in non-doped cavities.

Looking at the curve of Q-factor as a function of the accelerating field of such BCP 3.9 GHz cavities, we can clearly recognize the peculiar anti Q-slope, that had been previously observed only in N-doped cavities. 

Important implication of this new findings is that high frequency cavities might be useful for high field applications. For example, we have found that the 120 C baked 2.6 GHz cavity shows Q-factors that approaches the one of a 1.3 GHz cavity processed with same treatment between 35 and 40 MV/m.

\begin{figure}[t]
\centering
\includegraphics[scale=1]{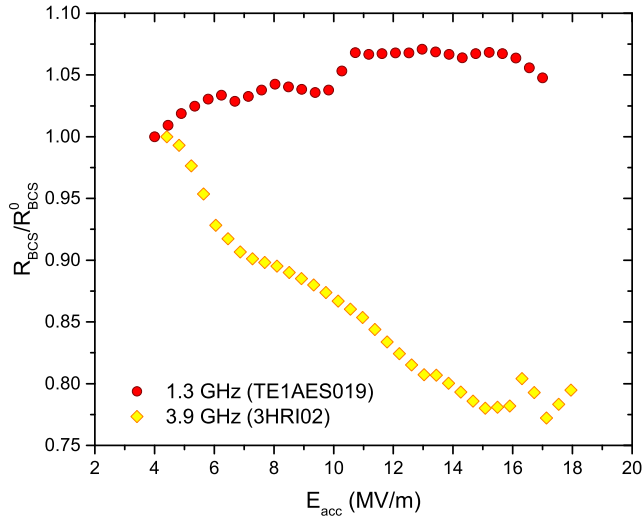}
\caption{Normalized data $ R_{BCS}/R_{BCS}^0 $ as a function of the accelerating field for  BCP cavities.}
\label{BCP}
\end{figure}

Interestingly enough, comparing $ R_{BCS}/R_{BCS}^0 $ for N-doped cavities (made with exact same doping recipe) at 650 MHz, 1.3 GHz, 2.6 GHz and 3.9 GHz (Figure 4), it is possible to notice similar results. At 650 MHz the BCS surface resistance of N-doped cavities slightly increase with the field, in contrast with the typical decreasing observed at 1.3 GHz. On the other hand, at 2.6 and 3.9 GHz, the effect of the reversal of $R_{BCS}(E_{acc})$ is enhanced. At 3.9 GHz, in particular, we can noticed that the BCS surface resistance is substantially decreased around $15-20$ MV/m, comparing with the value at low filed.

\begin{figure}[t]
\centering
\includegraphics[scale=1]{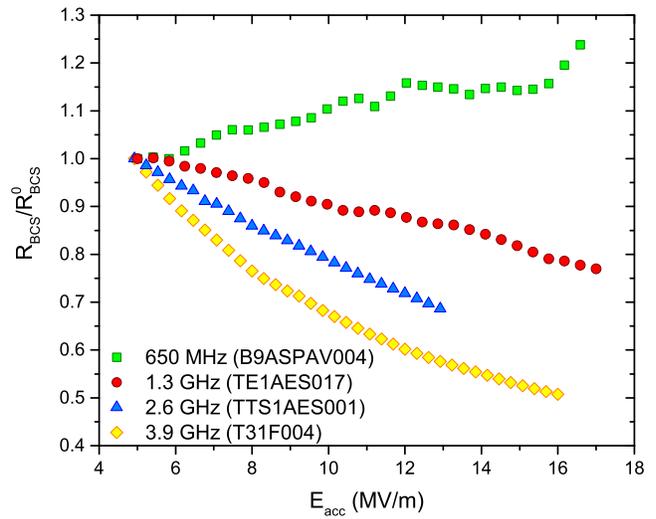}
\caption{Normalized data $ R_{BCS}/R_{BCS}^0 $ as a function of the accelerating field for  N-doped cavities.}
\label{Ndop}
\end{figure}

Also these findings reveal very important consequences. Thanks to such substantial decreasing of the BCS surface resistance, we observed unprecedented high value of Q-factors at medium field with the N-doped 3.9 GHz cavity, that reached $Q_0 \sim 1.5 \cdot 10^{10}$ at about 20 $MV/m$.

These field dependent trends of the BCS surface resistance represent an important suggestion that niobium SRF cavities may operate under a non-equilibrium regime depending on the resonance frequency and the types and concentration of impurities. In agreement with the Eliasberg theory \cite{Elias}, non-equilibrium effects in superconductors are visible only above a certain frequency threshold. This is due to the fact that the RF field period has to be shorter than the quasi-particle relaxation time in order to increase the population of quasiparticles at high-energy states far from the gap edge. 

Since the recombination into Cooper pairs is more favorable when the quasiparticles occupy higher energy states, by increasing their population far from the gap edge the recombination is enhanced and the number of Cooper pairs increases \cite{Scal}. Such a regime of stimulated superconductivity may imply the manifestation of the phenomenon of decreasing of the surface resistance as a function of the accelerating field, as experimentally observed for SRF cavities. 

\section{Conclusions}

From the data here presented it is clear that the physical mechanism underneath the reversal of the BCS surface resistance usually observed in N-doped cavities show a stronger effect at high frequencies. On the other hand, we have found that below certain frequency the effect of the reversal of the BCS surface resistance is not observed, even in N-doped cavities.

Very interesting we have observed for the first time the anti Q-slope effect in clean Niobium cavities resonating at 3.9 GHz. Decomposing the surface resistance in its different contributions, we were able to reveal that also in this case the anti Q-slope effect is due to the decreasing of the BCS surface resistance as a function of the field.

Concluding, the field dependence of the BCS surface resistance do not depend only on the presence of interstitial impurities but it also strongly depends on the frequency of the RF field. 

\section{Acknowledgments}

This work was supported by the United States Department of Energy, Offices of High Energy and Nuclear Physics. Fermilab is operated by Fermi Research Alliance, LLC under Contract No. DE-AC02-07CH11359 with the United States Department of Energy.

% this setting when the default (\flushend)
% => "balance two column" shows bad results
%

\end{document}